\documentclass[10pt]{article}
\usepackage[latin1]{inputenc}
\usepackage[english]{babel}
 
\usepackage{color}
\usepackage{latexsym}
\usepackage{amssymb}
\usepackage{amsfonts}
\usepackage{amsmath}
\usepackage{graphicx}
\def\be{\begin{equation}}
\def\ee{\end{equation}}
\def\bc{\begin{center}}
\def\ec{\end{center}}

\begin{document}

\author{Andrea Scharnhorst$^\dagger$, 
Eugene Garfield$^\ddagger$ \\ 
{\small $^\dagger$ Royal
    Netherlands Academy of Arts and Sciences (KNAW)}\\ {\small The Virtual
    Knowledge Studio for the Humanities and Social Sciences} \\ 
{\small $^\ddagger$
    Thomson Reuters Scientific (formerly ISI)} \\ 
{\small andrea.scharnhorst@vks.knaw.nl}}
 
\title{\bf Tracing scientific influence}

\maketitle
 
\footnote{Dynamics of Socio-Economic Systems, Vol. 2, Number 1: xx-xx (2010). ISSN 1852-379X. Received: 21/04/09, Accepted: 26/08/10. http://www.dyses.org.ar/IJ-DySES}

\begin{abstract}
Scientometrics is the field of quantitative studies of scholarly activity. It has been used for systematic studies of the fundamentals of scholarly practice as well as for evaluation purposes. Although advocated from the very beginning the use of scientometrics as an additional method for science history is still under explored. In this paper we show how a scientometric analysis can be used to shed light on the reception history of certain outstanding scholars. As a case, we look into citation patterns of a specific paper by the American sociologist Robert K. Merton. 
\end{abstract}

\section{Introduction}
Scientometrics is a field of information and communication sciences devoted to quantitative studies of science. The term was coined by V.V. Nalimov and Z.M. Mul'chenko as title of a book about the measurement of scientific activity. \cite{nalimov1969}
The systematic study of systems of thoughts was and still is inherent part of philosophy. But the growth of the academic system after World War II, the need for accountability of public spending, the increasing role of technological innovation for economic wealth, and critical debates about the role of science for society lead to a formation of a new special field devoted to the study of scholarly activity. In this newly emerging field the more traditional epistemic and historical perspective has been combined with studying the sciences as a social system using approaches from social-psychology, sociology, and cultural studies. Bernal \cite{bernal1939} has been called one of the grandfathers of this emerging field \cite{steiner1989}, and the foundation of a society called   \textquoteleft Society for Social Studies of Science\textquoteright\/ in 1975 \footnote{See http://www.4sonline.org/} was a first sign of an institutional consolidation of the scientific community interested in the   \textquoteleft sciences\textquoteright\/ as an object of studies. At the very beginning, quantitative studies and qualitative approaches were closely together.\cite{elkana1978} The sociological theories of Robert K. Merton about feedback mechanisms (social enforcement) in the distribution of award in the science system \cite{merton1968,merton1988,merton1995} resonated with stochastic mathematical models for the skew distribution of citations as proposed by Derek de Solla Price, a physicist and science historian. \cite{price1965,price1976} Dobrov and others proposed a socio-economic theory of the academic systems shedding light on necessary preconditions of scientific labor and related, so-called input indicators. \cite{dobrov1971} The emergence of (digital) databases of scientific information such as the \textit{Science Citation Index} of the \textit{Institute for Scientific Information} (ISI, now Thompson Reuters) \cite{cronin2000} triggered a wave of systematic, statistical studies of scientific activities - the core of \textit{scientometrics} still today.
\subsection{Texts and authors}
Not surprisingly, the number of quantitative studies grew with the availability of data. Most of the scientometric studies were devoted to products of scholarly activity, namely publications. They are based on a so-called \textquotedblleft literary model of science\textquotedblright\/   and have been boosted by the groundbreaking innovation of a bibliographic information system which includes the references used in a paper - on top of authors, title, abstract, keywords and the bibliographic reference itself.\cite{wouters1999} The last decades have witnessed a bias of quantitative studies about the products (texts and communication) of scholarly activity compared to studies of their producers (authors) or the circumstances of the production (expenditures). In Fig. 1 some relevant branches of research inside of scientometrics and some of their representatives are named. 

\begin{figure}[!h]
    \centering
    \includegraphics[width=12cm]{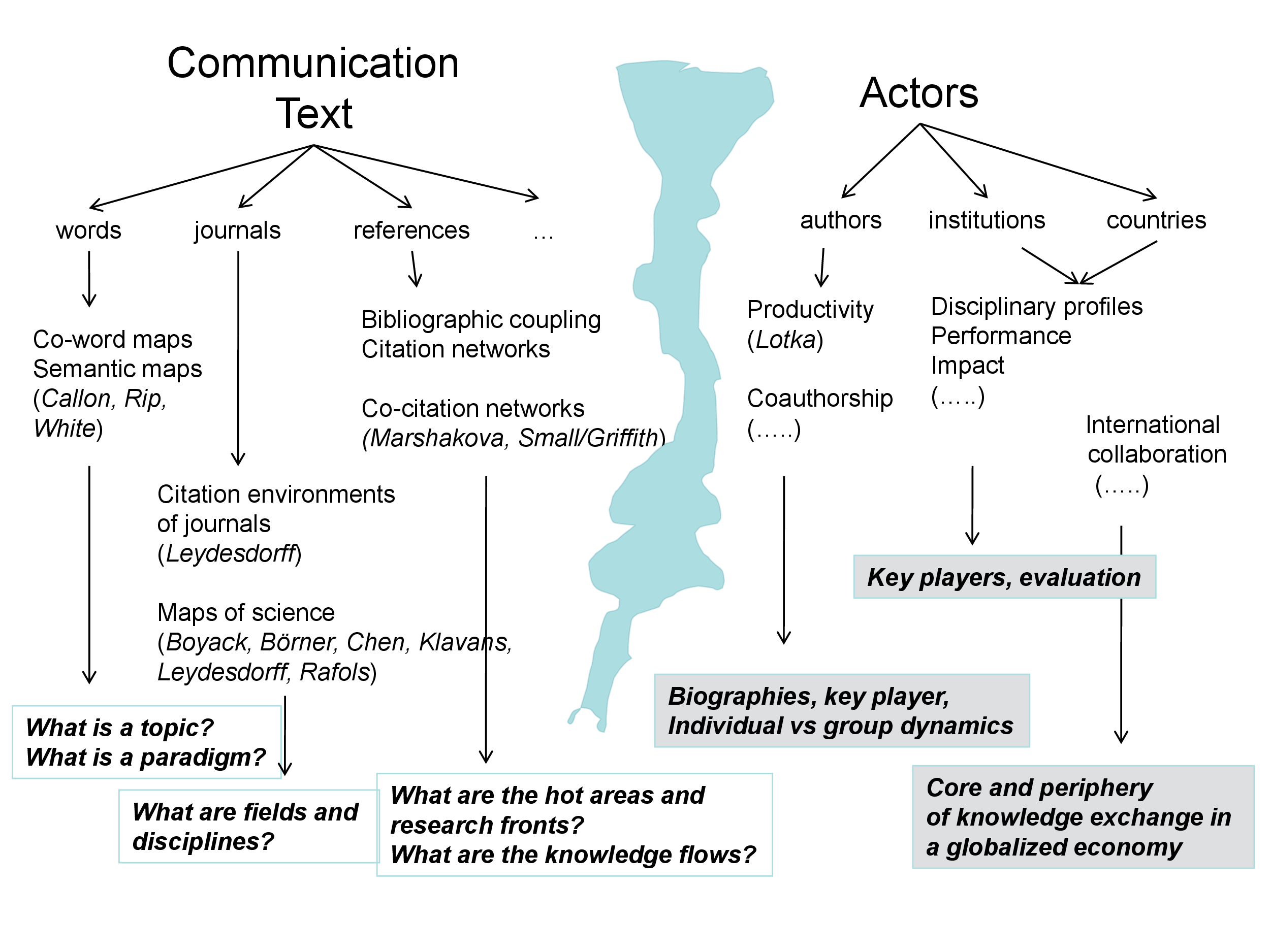}
    \caption{Objects of scientometrical and bibliometrical studies arranged by their main units of analysis: texts and authors}
    \label{andrea1}
\end{figure}

This illustration does not claim any completeness. For a more comprehensive introduction into scientometrics we would like to refer the reader to a recently published book on bibliometrics \cite{debellis2009} which also discusses the social theories used in scientometrics. Further useful sources are the lecture notes of Wolfgang Gl\"anzel devoted to the main mathematical approaches to scientometrics indicators \cite{glaenzel2003}, the website of one of the authors\footnote{See www.eugenegarfield.org. In particular, we recommend the use of the   \textquoteleft search function\textquoteright\/ which allows a full text search through all fifteen volumes of the \textit{Essays of an Information Scientist}.} and, of course, main journals in the field such as \textit{Scientometrics}, \textit{Journal of the American Society for Information Science and Technology}, \textit{Journal of Documentation}, \textit{Research Policy}, \textit{Research Evaluation}, and \textit{Journal of Informetrics}.

The representation in Fig. \ref{andrea1} suggests that for large parts of scientometrics authors were the \textquotedblleft forgotten\textquotedblright   units of analysis. There is indeed a rationale behind the focus on texts. For the elements of textual production - or more specific journal articles - databases as the Citation Indices of the ISI - Web of Knowledge have introduced standards for the units of a bibliographic reference: the journal names, the subject classifications, and other meta data such as document type. However, the identification of authors by their names creates a problem (i.e., occurrence of common names, transcription of non-English names, name changes). Only recently attempts have been made to also make authors automatically traceable. One way is to introduce standardized meta data for authors, for instance by introducing a unique digital identity for reserchers. Currently, different systems -- commercial and public -- coexist and compete. Publishers as Thompson Reuters (see www.researcherid.com) and Elsevier have introduced ID's for authors. For the Dutch national science system the \textit{SURF Foundation} has introduced a Digital Author Identifier (DAI). Open repositories also aim for the identification of authors (see http://arxiv.org/help/author\_identifiers). Another way is to automatically allocate articles to authors using author-specific characteristics or patterns (e.g., a combination of a specific journal set, subject categories, and addresses). 

For large scale statistical analyses of the behavior of authors the ambiguity of person names is less important. Examples are investigations of authorship networks \cite{newman2004}, author's productivity \cite{fronczak2007}, or author-citation network models \cite{boerner2004}. But without a researcher identity, or additional knowledge about the author it is usually not possible to trace individual actors. The above mentioned steps, including a ResearcherID, might allow more systematic author based studies beyond samples sizes which still can be cleaned by manual inspection. In the future a combination of following the creators of scientific ideas and the influence of these ideas themselves seems to be possible and promising.\cite{leydesdorff2010}

However, the dichotomy between texts and authors as introduced in Fig. \ref{andrea1} is not a strict one. Scientometric studies cannot be always fully separated into either text and author centered. There is a gray area between both directions and interesting studies can be found -- also in the history of scientometrics -- which trace authors in threads of ideas and \textit{vice versa}. Two of them have inspired this paper: \textbf{algorithmic historiography} and \textbf{field mobility}. 
\subsection{Algorithmic historiography and field mobility}
\begin{figure}[!h]
    \centering
    \includegraphics[width=10cm]{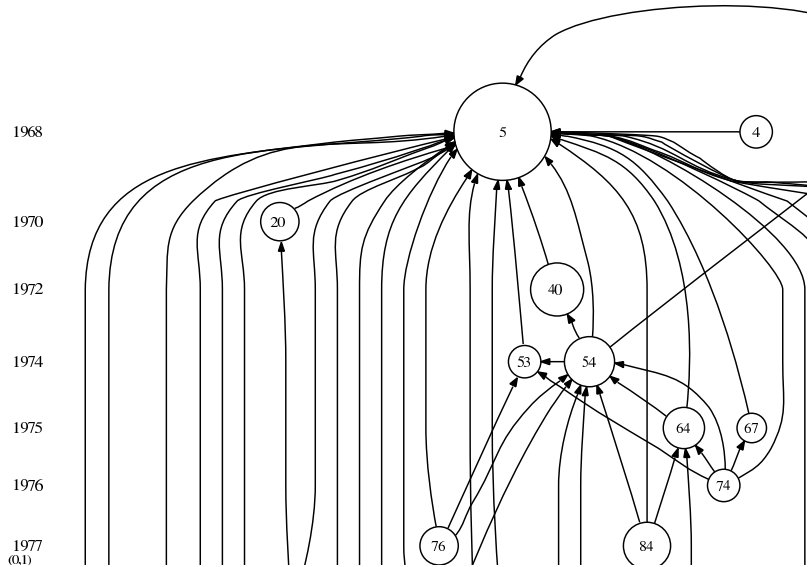}
    \caption{Historiograph of Merton's paper of 1968 \cite{merton1968} --
      Snapshot from a part of the online available graph G2\hfill\ \newline
 (http://garfield.library.upenn.edu/histcomp/merton-matthew-I/index-tl.html). The nodes represent selected papers citing Merton's paper. Node 5 on top is Merton's own original paper \cite{merton1968}. The size of a node (paper) is proportional to the number of in-coming links (citations).}
    \label{andrea2}
\end{figure}

The so-called \textbf{historiographic approach} has been proposed by the second author of the present article. This approach allows to reconstruct the main citation paths over years starting from a seed node. The seed can be one paper of an author, or a collection of papers characterizing an author or a scientific speciality.  Based on ISI-data, the tool \textit{HistCite} allows to extract and to visualize the citation network.\cite{garfield2003} Based on the citation rates of all nodes in this network either in the network itself (local -- G1 graph) or in the whole ISI database (global -- G2 graph) graphs can be displayed showing the citation tree of the most cited papers in this directed graph ordered along a time axis.\footnote{The description of the tool and an extended number of cases results are available on-line, see  http://garfield.library.upenn.edu/histcomp/guide.html.}  Through a historiographical analysis one can reconstruct schools of thoughts; paths of influence and the diffusion of ideas.\cite{lucio2008} In this paper, we use one of the recorded \textit{HistCite} files, namely the historiograph of Merton's paper of 1968 \cite{merton1968}. A part of the global \textit{HistCite} graph G2 is displayed in Fig. \ref{andrea2}.

A novel aspect applied in the present paper is an analysis of the nodes in the historiograph in terms of their disciplinary origin. This approach has been inspired by the study of \textbf{field mobility} -- a term coined by Jan Vlachy.\cite{vlachy1981} Vlachy introduced this notion of mobility, generalizing geographic and occupational mobility or migration towards the movement of researchers through cognitive spaces. \textit{Field mobility} describes one aspect of the \textit{cognitive mobility} of a researcher who during the life span of her or his career moves from scientific field to scientific field. In the past, this approach has been integrated into a dynamic model of scientific growth.\cite{bruckner1990} Recently, the concept has been used again to trace the activities of a researcher in different fields.\cite{hellsten2008} The crucial point for an application of this concept is the question how to identify the fields. For the traces of individual authors mobility Hellsten et al. \cite{hellsten2008} used self-citations. Self-citations represent a self-referential mechanism which automatically leads to a clustering of papers which have a common focus. Often in these thematic clusters we also find different co-authors, and different keywords and title words can be used to label the different \textit{fields of activity} on the micro-level of an individual researcher. 

The analysis we present in this paper is a combination of a historiographic and a field mobility approach. We extend the approach of field mobility from the mobility of a researcher between fields to the mobility of a paper between fields. While one can easily imagine that a researcher by her or his on-going creativity travels between topics and fields, this is intuitively less clear for a published -- and therefore stable -- paper. So, what do we mean by this? Once a paper is published, it has a certain location in an envisioned landscape of science. This position can be determined by the journal in which the paper appeared. The disciplinary classification of the journal can be seen as an attribute or characteristics of the paper which allows to place it on a map of science.\cite{boyack2005} While this landscape seems to be relatively stable or, at least, slowly changes over time, citations to a paper represent a more fluid and faster dynamics. A paper published in sociology, for instance, can suddenly gain importance for different areas as distant as physics or computer sciences. If we look at this paper through the lense of citations its position can be variable. Eventually, the changing perception of the paper causes its travel in this imagined landscape. Referencing to papers can be seen as a process of re-shaping the scientific landscape. Due to sequential layers of perception the actual   \textquoteleft location\textquoteright\/ of a paper, now determined by the position of the recent papers citing it, can shift. This travel of a paper, or more precisely its perception, between fields is an indicator for the diffusion of ideas and \textit{field mobility} in a generalized sense. In difference to the earlier mentioned author mobility study \cite{hellsten2008} in this paper we determine different fields by manually inspecting and classifying journals. This way we identify \textit{fields of activity} on the meso-level of journals, rather than on the micro-level of individual behavior (as done in the case study on self-citation pattern), or on the macro-level of disciplines (as represented by larger journal groups).  
\subsection{The Matthew effect of science -- Merton's famous paper of 1968}
Due to its broad and persistent perception beyond sociology, Merton's work seems to be a good candidate for studying the diffusion of ideas. Moreover, looking at citation behavior over time and mobility phenomena we want to shed light on the micro-dynamic processes at the basis of past, current and future structures of the landscape of science. In a certain sense, we study the same questions as Merton who himself asked for generic mechanisms in the dynamics of science. In the paper under study -- his 1968 paper on the \textit{Matthew effect of science} \cite{merton1968} -- he proposes a specific mechanisms, namely the accumulation of reward and attention. With our study, we ask to which extend we can use the dynamics of the perception of his work as a case to shed another light on these generic mechanisms. 

The influence and relevance of Merton's work has been discussed earlier by one of the authors.\cite{garfield1973, garfield2004a, garfield2004b} Historiographs of his \OE uvre or part of it are available for further inspection.\footnote{See www.eugenegarfield.org} Still, it remains a question what actually bibliometrics can add to science history based on text analysis and eye witness accounts. Recently, Harriet Zuckerman \cite{zuckerman2010} has thoroughly discussed the Matthew effect and carefully analyzed its perception in past and presence. In her analysis she uses bibliometric information for the global pattern of perception of Merton in a kind of \textit{bird's-eye view}. The current bibliometric exercise complements her study on a meso-level. In this paper, we concentrate on the perception of one specific paper of Robert K. Merton \cite{merton1968} out of the three devoted to the \textit{Matthew effect} \cite{merton1968,merton1988,merton1995}. Instead of looking on the overall citation numbers (macro-level) or following the nodes and paths in the historiograph of this particular paper in depth (micro-level), we analyze the citing papers according to the disciplinary distribution of the journals in which these papers appeared (meso-level). 
\section{Merton's legacy -- the surprising longevity of his paper of 1968}
Robert K. Merton (1910-2003) is known for his theory of social structures as an organized set of social relationships, the discussion of their functionality or disfunctionality, and for his definition of culture as an \textquotedblleft organized set of normative values governing behavior\textquotedblright\/.\cite{merton1949} Applied to science(s) as a social system, he defined four scientific norms or ideals: communalism, universalism, disinterestedness, and organized skepticism. Looking for empirical evidence supporting or undermining theoretical frames, he was also interested in social behavior of scientists which actually contradicts the norms and values functional for science.\cite{merton1957} In particular, he drew attention to mechanisms of reward in science. In 1968 he published a paper in the journal \textit{Science} entitled \textquotedblleft The Matthew Effect in Science: The Reward and Communication Systems of Science are Considered\textquotedblright\/ \cite{merton1968}. In this paper he addressed the phenomenon that well-known scientists often receive more reward for the same contribution than other, less-known researchers.

While this \textquotedblleft the rich get's richer effect\textquotedblright\/  has often been described as a sign of injustice and malfunctioning, Merton also discussed that this deviant behavior from an ideal one has a constitutive, positive function for the whole system. It creates a focus of attention, a kind of pre-selection, and a structuring which allows an easier orientation in large amounts of information. 
\subsection{The Matthew effect of science -- an example of a positive feedback mechanism}

It has been argued elsewhere that the essence of the \textit{Parable of the Talents} as told in the Bible, does not so much concern an unequal distribution of wealth or reward as such, but the difference between an expected and eventually achieved position in such a rank distribution depending on an appropriate \textit{use of talents}.\cite{bonitz1997} In an empirical study of the scientific performance of countries in terms of citation gathering it has been shown \cite{bonitz1999} that \textquotedblleft privileged countries\textquotedblright\/  in terms of expected citation rates receive even more citations than countries with smaller expectations. Not only is the distribution of talents, gifts, strengths a skewed one, their further use seems to even increase this skewness.

Apparently, the \textit{Matthew effect} does not play in favor of certain researchers and allocates fame and reward not always to the person which deserves it most. However, on the level of the system, this effect is an important dynamic mechanism to create order out of chaos.\cite{prigogine1986} Different authors -- among them Derek de Solla Price \cite{price1976} -- have pointed to the fact that in the language of system theory, cybernetics, and mathematics, this effect corresponds to a positive feedback loop which introduces a non-linear interaction mechanisms into the dynamics of the system. In terms of mathematical models this can be described as self-accelerating growth rate, a growth rate of an entity depending on the actual size of the entity itself. Thereby, an entity can be a scientific field, a certain technology, or a certain type of behavior. Applied to a single entity models with such growth rates describe non-linear growth, exponential or hyperbolic. When implemented as a mechanism in a system of several competing growing entities different types of selection, including hyperselection, can result from this non-linear mechanism.\cite{bruckner1996} In general, positive feedback loops -- such as the \textit{Matthew effect} -- are at the core of specific pattern formation visible in skew distributions \cite{adamic1999} or in dominant designs \cite{suarez1995}. It is therefore not surprising that the perception of Merton's paper \cite{merton1968} has not been restricted to sociology and other social sciences, but has also found resonance in mathematics and physics.

\subsection{Citation pattern}

We started our investigation with the question why Merton's paper \cite{merton1968} still so often appears in the list of references of various authors. Citation numbers of Merton's paper seem even to show an increase instead of an expected fading away. This bibliometric observation seems to be in line with other observations of contemporary witnesses and friends of Merton. Our question was: Can we use scientometrics to find objective, data-based evidence for subjective impressions? Is there any way to \textit{factualize} the impact of this specific paper of Merton?

If we explore what citation analysis can contribute to uncover some attributes of the lasting impact of Merton's work, we face the analytic challenge to measure the impact of a single work (a single paper) with methods designed to reveal regularities in large amounts of data. Let us therefore first present some standard scientometrical insights into the citation history of scientific publications.

Citation analysis has taught us about so-called \textit{Citation Classics}. These are highly cited papers -- sometimes even Nobel prize winning ones. Beginning in the 1960's, one of the authors started to publish about highly cited papers in the journal \textit{Current Contents}. Highly cited papers represent only a very small fraction of all papers and citation rates are highly field-dependent. Being aware of this, a refined methodology was proposed to identify a \textit{Citation Classics}. \textquotedblleft As few as 100 citations (or even fewer) may qualify a work as a Citation Classic in some of these areas, such as radio astronomy, engineering, or mathematics. To identify Citation Classics in smaller fields we use several criteria. One is rank within a specialty journal. If a specialty journal defines a unique field, then the most-cited articles from that journal include many if not all Citation Classics for that field.\textquotedblright\/ \cite{garfield1986}

Merton as an author has entered the set of \textit{Citation Classics} not with his paper of 1968 \cite{merton1968} but with his book \textit{Social Theory and Social Structure} from 1949 \cite{merton1949}. In his own commentary on this fact Merton wrote: \textquotedblleft I am not at all sure of the reasons for Social Theory and Social Structure (STSS) still being cited 30 years after its first appearance. To answer that question with reasonable assurance would require a detailed citation analysis and readership study, hardly worth the effort.\textquotedblright\/ \cite{merton1980} Nowadays, in the age of digitally available databases and computers, such an effort is more practical. 

\begin{figure}[!h]
    \centering
    \includegraphics[width=10cm]{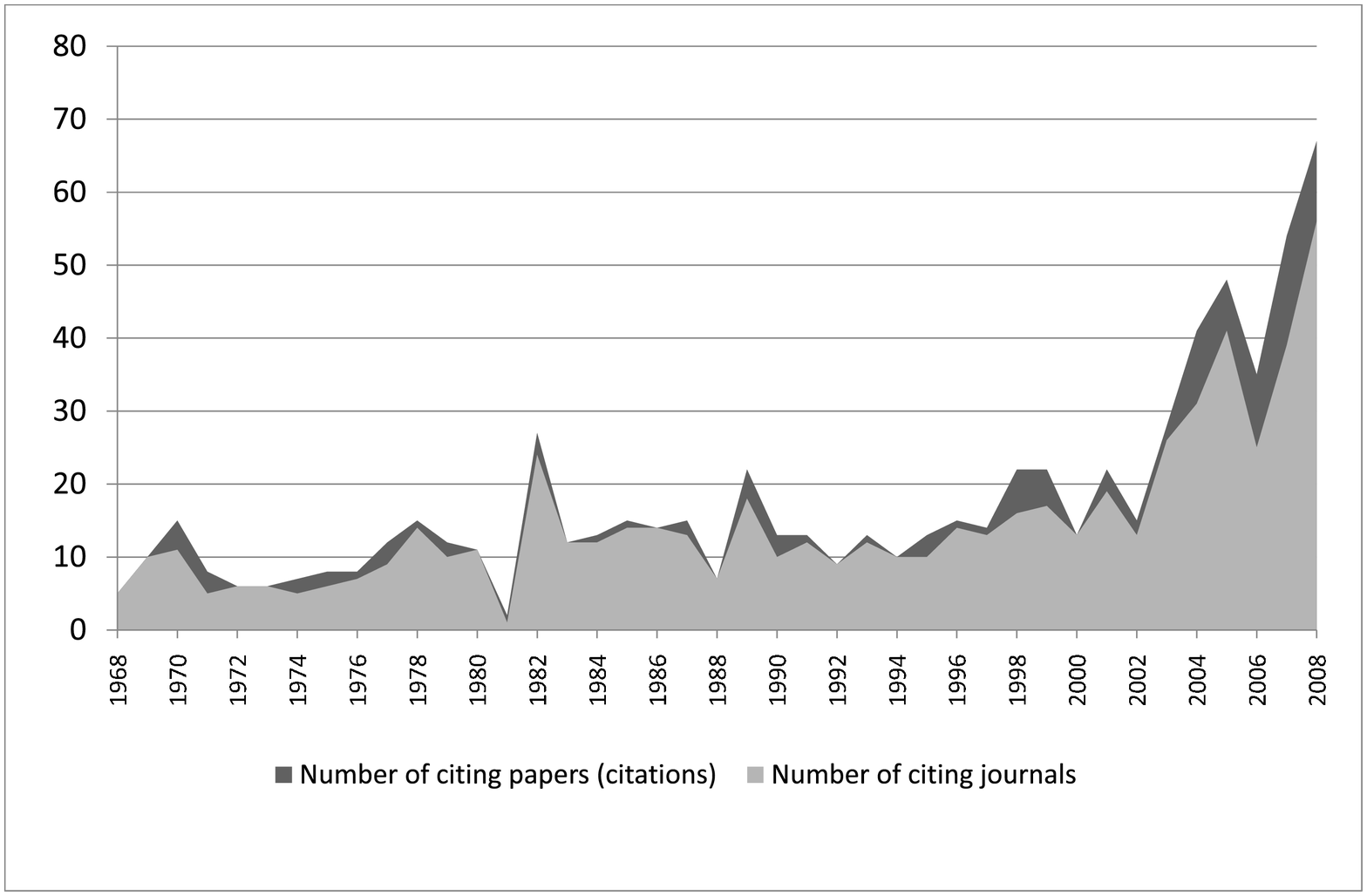}
    \caption{Annual number of citations to Merton's paper \cite{merton1968} from 1968 to 2008 (citing papers (dark) and citing journals (light))}
    \label{andrea3}
\end{figure}

So, let us first look at the citation number for Merton's paper of 1968 \cite{merton1968} (see Fig. \ref{andrea3}).
Merton's paper has attracted 741 citations from 1968 to June 2009\footnote{See also http://garfield.library.upenn.edu/histcomp/merton-matthew-I/index-tl.html for the raw data. Note that the set of citations used for the on-line available HistCite files and the data set used for this analysis differ for 3 articles.}. To extract these data we used the \textquotedblleft cited reference search command\textquotedblright\/ in the \textit{Web of Knowledge}\footnote{See http://isiknowledge.com/.}. Records of the retrieved citing documents have been downloaded and exported into an EXCEL data base for further analysis. 
The annual number of citing publications fluctuates between 5 and 15-20 over the period of more than 30 years with a slightly increasing tendency, but from 2002 onwards we observe a remarkable increase of it (see Fig. \ref{andrea3}). Even if one takes into account that the database itself is not steady but growing over time, we can state that the perception of Merton's work is continuing. 

From Fig. \ref{andrea3} we can also see that the citations to Merton's paper \cite{merton1968} are widely scattered. In the figure we display the number of citing papers and the number of citing journals together. Whenever in Fig. \ref{andrea3} both graphs coincide each of the citations appear in a different journal. Whenever the dark gray area is seen above the light one, in some journals more than one paper cites Merton \cite{merton1968}. We will look into these multiple citations from journals later again. A closer look reveals that not only the number of citations increases but also the number of journals with papers citing Merton. 

Citation analysis also taught us about different possible citation life-cycles of a paper, a person, or a research field. Vlachy developed a typology of these life courses of papers in terms of citations. Successive citations represent traces a work leaves in our collective memory. We see patterns between \textquotedblleft never reaching a wider audience\textquotedblright\/  (\textquoteleft scarcely reflected\textquoteright\/ as labeled by Vlachy), \textquotedblleft oscillating recognition\textquotedblright\/, \textquotedblleft exponential or hyberbolic growth\textquotedblright\/  (\textquoteleft genial\textquoteright\/), and an almost \textquotedblleft Gaussian\textquotedblright\/ growth and decline of recognition (\textquoteleft basic recognized\textquoteright\/).\cite{vlachy1983}

The latter effect seems to be much in line with what the analysis of larger ensembles has shown, namely that there is a citation window and a \textquotedblleft half-life\textquotedblright\-time of a paper. Moreover, after a certain peak in recognition the knowledge related to a certain paper becomes incorporated into reviews, textbooks, or figures under the name of an effect or author only without carrying a citation mark anymore.

But, patterns and laws in the collective production of scientific knowledge is only one side of the coin. Important singular events -- critical (re)shaping the way we think about problems and solve them -- are another. Both sides do not contradict each other. Even more, they heavily rely on each other. For Merton's paper of 1968 \cite{merton1968} we find a steady growth over decades. What causes this growth? To answer this question we analyze the journals which contain the citing papers.

\subsection{Journal traces}
\subsubsection{Distribution over journals}

\begin{figure}[!h]
\centering
\includegraphics[width=12cm]{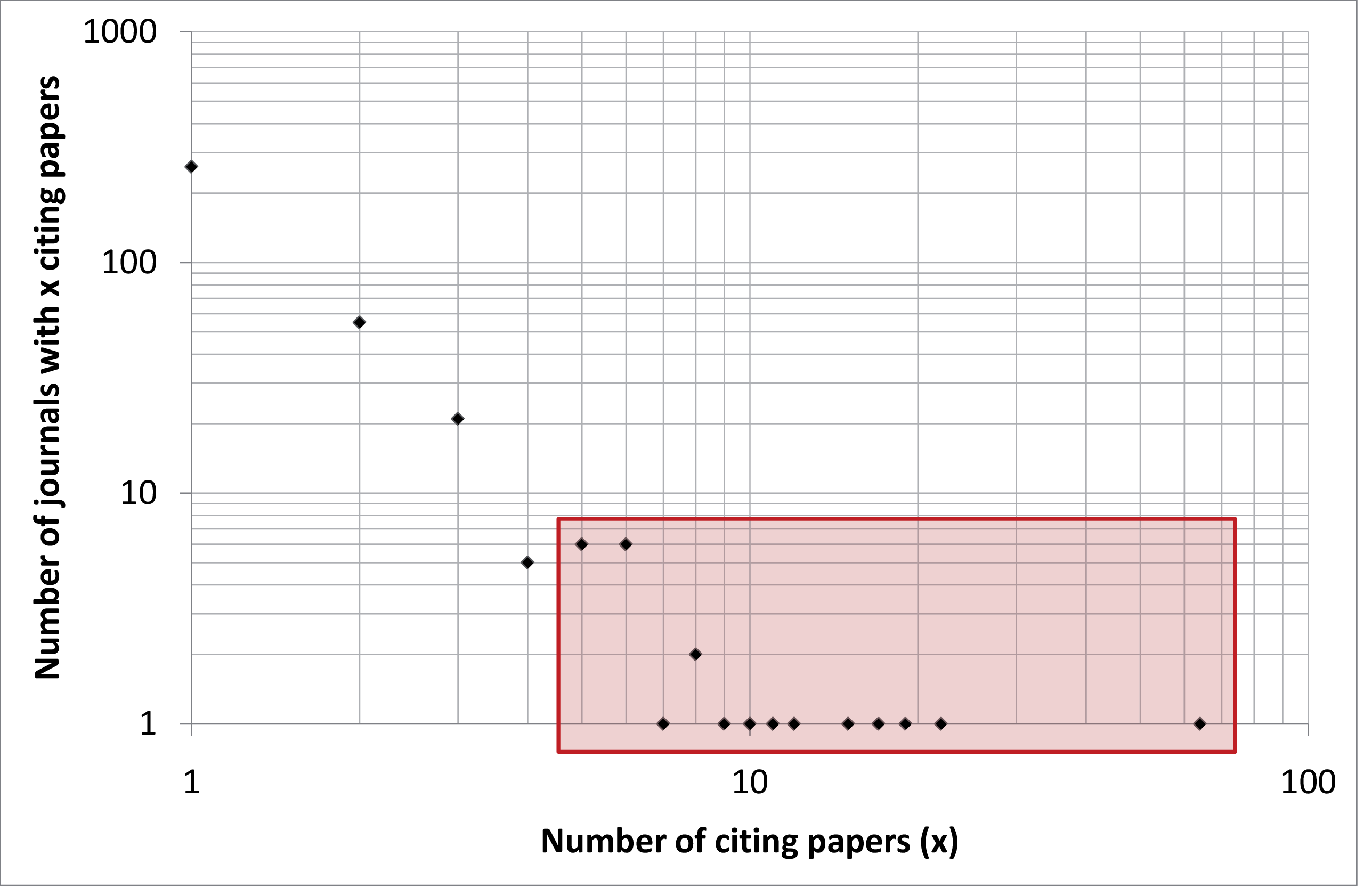}
\caption{Plot of journals against citing papers. Core set marked by a grey rectangular.}
\label{andrea5}
\end{figure}

The citations towards Merton's paper \cite{merton1968} are concentrated in some of the 368 journals over which the citations in the whole period are distributed.  
If we plot the number of journals with n citations against n we see that only 24 journals carry more than 5 citations in the whole period (Fig. \ref{andrea5}). 
%
%

\subsubsection{Classification of journals}

In the next step we allocated the journals to fields using two different classifications. A finer classification was used for a core set of 24 journals with more than 5 citations (see Table \ref{table1}). This core set carries about 40 percent of all citations. For the further examination of the whole journal set we used a rougher classification on the level of disciplines (see Table \ref{table2}). We choose field and discipline names used in bibliometric studies. The allocation of a journal to it is based on personal judgment. In both tables we also give the overall number of citations from these fields to Merton's paper \cite{merton1968}.

\begin{table}[!h]
\centering
\caption{Classification of the core journal group}
\begin{tabular}{|p{1cm}|p{7cm}|{c}|}
\hline
 & Field & Citing papers\\
 \hline
STI & Science and Technology Studies -- Information Science & 106\\
\hline
SOC & Sociology & 62\\
\hline
STE & Science and Technology Studies -- Evaluation & 28\\
\hline
EDU & Education & 25\\
\hline
STC & Science and Technology Studies -- Science Studies & 23\\
\hline
ID & Information and Documentation & 14\\
\hline
PSY & Psychology & 6\\
\hline
PHI & Philosophy & 5\\
\hline
MAN & Management & 3\\
\hline
\end{tabular}
\label{table1}
\end{table}

\begin{table}[!h]
\centering
\caption{Classification of the whole journal set}
\begin{tabular}{|p{1cm}|p{7cm}|{c}|}
\hline
 & Field & Citing papers\\
\hline
MATH & Mathematics & 2\\
\hline
PHYS & Physics & 21 \\
\hline
CHEM & Chemistry & 3 \\
\hline
ENG	& Engineering & 12\\
\hline
LIF	& Life Sciences & 20\\
\hline
MED	& Medical Research (including Psychology) & 94\\
\hline
SOC &	Sociology/Social Sciences/Information Science & 540\\
\hline
PHI	& Philosophy & 30\\
\hline
MULT & Multidisciplinary & 19\\
\hline
\end{tabular}
\label{table2}
\end{table}

\subsubsection{Analysis of the core journal set -- the field of Science and Technology Studies}

First, we took a closer look at the core set of journals. We asked which journals are part of this core set, which fields they represent, and how their presence in the core set changes over time. The growth of perception of Merton's paper \cite{merton1968} appears mainly in this core group of journals. In Table \ref{table3} we display the distribution of citing papers across journals for all journals with more than 5 citing papers. Merton's paper \cite{merton1968} was published in an interdisciplinary journal. It was first taken up in established sociological journals. But, most of the journals in the core set have only be founded in the 1960's or 1970's. In Table \ref{table3} we indicate the first year when a citing paper appeared, the category of the journal (using the classification given in Table \ref{table1}), and the founding date\footnote{Information according to major library catalogues such as of the \textit{Staatsbibliothek zu Berlin - Preu\ss ischer Kulturbesitz}, or of the \textit{British Library}.} of the journal. A comparison of the year of foundation of a journal and its first appearance in the core set shows that the perception of Merton's paper \cite{merton1968} co-evolves with the newly emerging field of \textit{Science and Technology Studies} (\textbf{ST*}). If we order the journals according to their overall number of citations, the journal SCIENTOMETRICS contains most of the papers citing Merton's paper of 1968 \cite{merton1968}. SCIENTOMETRICS is not the first journal in which Merton's paper is cited. Papers in the journals AMERICAN JOURNAL OF SOCIOLOGY, SOCIOLOGY OF EDUCATION, ANNUAL REVIEW OF INFORMATION SCIENCE AND TECHNOLOGY, AMERICAN SOCIOLOGICAL REVIEW and ACTA CIENTIFICA VENEZOLANA deliver the first 4 citations in 1968. But, the example of SCIENTOMETRICS represents the consolidation of a new field -- the field of \textit{Science and Technology Studies} which grows partly inside existing journals and partly due to newly emerging journals. About 30 percent of the journals in the core set belong to this field (\textbf{ST-I/E/C}) and contain about 60 percent of all citing papers.

\begin{table}
\centering
\caption{Distribution of citing papers among journals}
\begin{tabular}{|p{0.8cm}|p{0.8cm}|p{10.cm}|}
\hline 
\textsc{\# of citing papers} 
&  \textsc{\# of journals} 
& \textsc{Journal name (founding date) -- Field -- Year of the first citation of Merton's paper \cite{merton1968}}\\
\hline 
64 & 1 & SCIENTOMETRICS (1978)-- \textbf{STI} -- 1979 \\
\hline
22 & 1 & JOURNAL OF THE AMERICAN SOCIETY FOR INFORMATION SCIENCE AND TECHNOLOGY (2000)/JOURNAL OF THE AMERICAN SOCIETY FOR INFORMATION SCIENCE (1970, formerly AMERICAN JOURNAL OF DOCUMENTATION 1950)-- JASIST-- \textbf{STI} -- 1973 \\
\hline
19	& 1	& RESEARCH POLICY (1971) -- \textbf{STE} -- 1978\\
\hline
17 &	1	& SOCIAL STUDIES OF SCIENCE (1975, formerly Science Studies 1971) -- \textbf{STC} -- 1975\\
\hline
15 &	1	& AMERICAN SOCIOLOGICAL REVIEW (1936) -- \textbf{SOC} -- 1968\\
\hline
12 &	1	& CURRENT CONTENTS (1958)-- \textbf{ID} -- 1977\\
\hline
11	& 1	& AMERICAN JOURNAL OF SOCIOLOGY (1895) -- \textbf{SOC} -- 1968\\
\hline
10 & 1	& ANNUAL REVIEW OF INFORMATION SCIENCE AND TECHNOLOGY (1966) -- \textbf{STI} -- 1968\\
\hline 
9 &	1	& KOLNER ZEITSCHRIFT FUR SOZIOLOGIE UND SOZIALPSYCHOLOGIE (1948) -- \textbf{SOC} -- 1971\\
\hline
8 & 2	& AMERICAN PSYCHOLOGIST (1946) -- \textbf{PSY} -- 1969 \\ 
 & & HIGHER EDUCATION (1972) -- \textbf{EDU} -- 1988 \\ 
\hline
7	& 1	& ADMINISTRATIVE SCIENCE QUARTERLY (1956) -- \textbf{MAN} -- 1989\\ 
\hline
6	& 6	& 
SOCIOLOGY OF EDUCATION (1963) -- \textbf{EDU/SOC} -- 1968\\ 
 & & SOCIOLOGICAL INQUIRY (1961) -- \textbf{SOC} -- 1970 \\
 & & SOCIAL SCIENCE INFORMATION/INFORMATION SUR LES SCIENCES SOCIALES (1962) -- \textbf{STC} -- 1973 \\
 & & SOCIAL FORCES (1925) -- \textbf{SOC} -- 1989 \\
 & & RESEARCH IN HIGHER EDUCATION (1973)-- \textbf{EDU} -- 1989 \\
 & & RESEARCH EVALUATION (1991) -- \textbf{STE} -- 2003 \\
\hline
5 & 6 & SOCIAL SCIENCE QUARTERLY (1968) -- \textbf{SOC} -- 1982 \\
 & & JOURNAL OF HEALTH AND SOCIAL BEHAVIOR (1967)-- \textbf{SOC} -- 1977 \\
 & & JOURNAL OF HIGHER EDUCATION (1930) -- \textbf{EDU} -- 1977 \\
 & & MINERVA (1962) -- \textbf{PHI} -- 1985 \\
 & & CREATIVITY RESEARCH JOURNAL (1988) -- \textbf{PSY}-- 1995 \\
 & & ORGANIZATION SCIENCE (1990) -- \textbf{SOC} -- 2000 \\
\hline
4 & 5 & ... \\
\hline
3 & 21 & ... \\
\hline
2 & 55 & ...\\
\hline
1 & 260 & ...\\
\hline
\end{tabular}
\label{table3}
\end{table}

We classified the field of \textit{Science and Technology Studies} into subfields such as \textquotedblleft Information Science\textquotedblright\/  (general laws, quantitatively and mathematically oriented), \textquotedblleft Evaluation\textquotedblright\/  (indicator research), and \textquotedblleft Science studies\textquotedblright\/  (general laws, qualitatively oriented), and allocated journals to these fields (see Tables \ref{table1}, \ref{table3}). We are aware of the possible objection that such an allocation contains an element of arbitrariness. We also did not take into account that the profiles of the journals and their function for the scientific community partly overlap. Moreover, these profiles change over time. Also, the actual articles citing Merton's paper \cite{merton1968} might contentwise represent a different approach than expressed in the rough journal categorization. Despite these shortcomings the display of different fields in the \textquotedblleft recognition sphere\textquotedblright\/ (set of citing papers)  of Merton's paper \cite{merton1968} both for the core set and for the whole set reveals interesting insights.

If we examine the whole set of journals across the time scale we make two observations. Looking at the entry and exit of journals into the area of perception of Merton's paper \cite{merton1968} we see that most of the journals are \textit{transient} -- they appear only a few times in the \textit{recognition} or \textit{perception sphere} around Merton's paper \cite{merton1968}. This is why the core of journals containing citations is relatively small. Second, the distribution of citations across fields and disciplines does not remain stable over time.
\begin{figure}[!h]
    \centering
    \includegraphics[width=10cm]{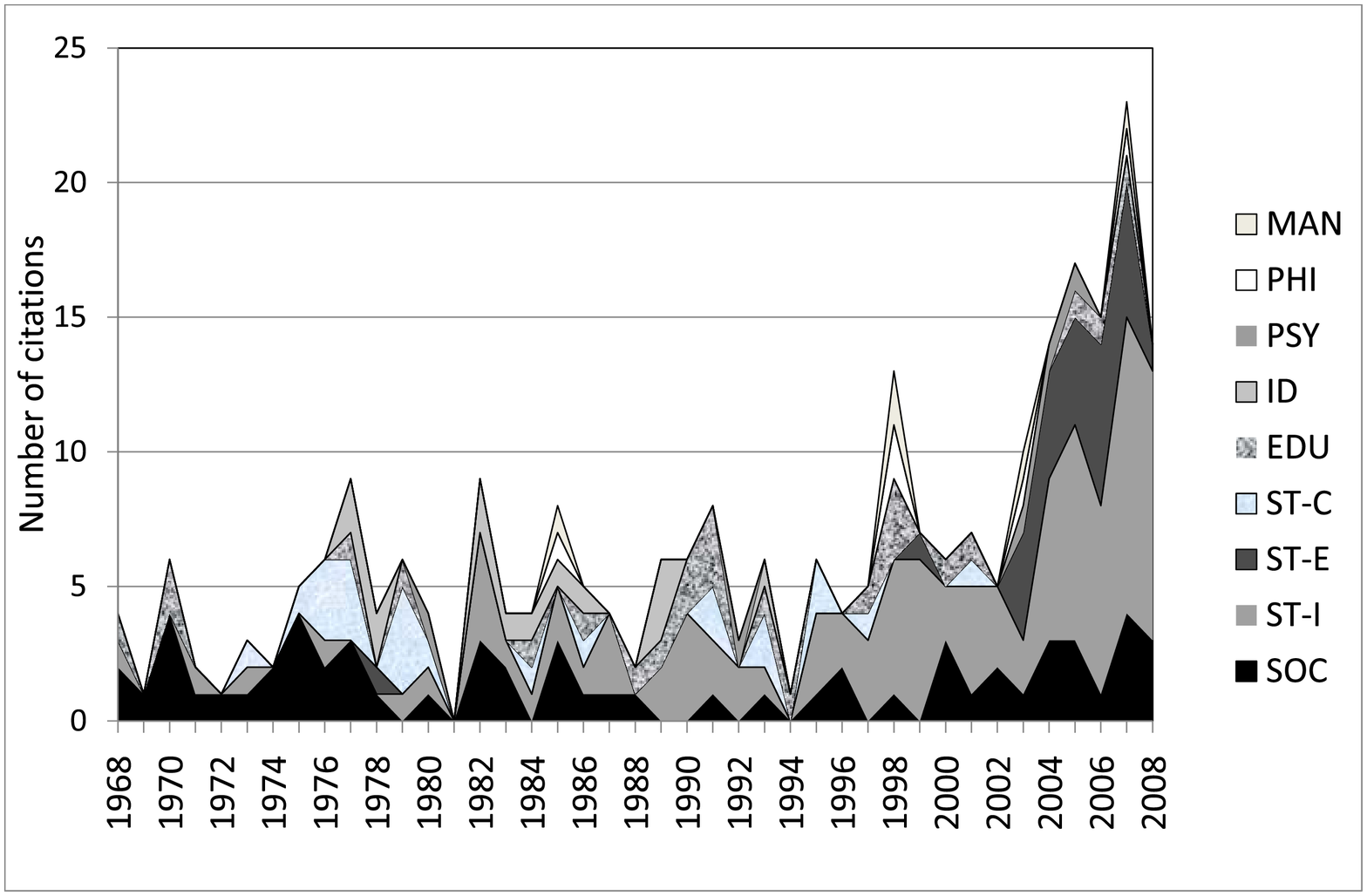}
    \caption{Disciplinary origin of the papers citing Merton \cite{merton1968} in the core set over time. For 1981 there is no citation in the core group.}
    \label{andrea6}
\end{figure}

In Fig. \ref{andrea6} we plot the number of citing papers per journals of the core set against the time axis. Next to an absolute increase we also observe a shift of attention among different fields. Not surprisingly the perception of Merton's paper \cite{merton1968} starts in sociological journals. For instance, in 1970 the journals AMERICAN JOURNAL OF SOCIOLOGY and SOCIOLOGICAL INQUIRY contribute with two citations each. \textit{Sociology} remains a persistent discipline over the years. The eventually dominating field of \textit{Science and Technologies Studies} gains momentum since mid of the 1970's. For instance, in 1977 the journal SOCIAL STUDIES OF SCIENCE contributes with three papers. In later years, in particular since the end of the 1990's \textbf{ST*} fields contribute with around 10 citations per year. 

If we look into the subfields of \textbf{ST*}, we see a clear shift from sociological and cultural studies of science towards informetric analysis. However, one has to take into account that all these statements are based on rather small numbers and, therefore, are susceptible to random factors. Only a close reading of the text of the citing papers could reveal if also the context of the citation to Merton's paper \cite{merton1968} changed systematically. By a random inspection we find papers reporting \textit{personal experiences} explained with the \textit{Matthew effect}, discussions of the social function of the effect, or its possible quantitative validation. 

In the core journal set, \textit{Philosophy} plays a rather marginal role. But, this does not mean that authors of philosophical journals are not interested in Merton's work. On the contrary, an analysis of the whole journal set shows that the discipline \textit{Philosophy} holds the third rank (see Table \ref{table2}). The explanation can be found in the wide scattering of citing papers over journals. In the core journal set only the journal MINERVA represents the field of \textit{Philosophy} with 5 citing papers. In the whole journal set we find 20 more journals classified under \textit{Philosophy}. Most of them appear only once. To detect the field-specific pattern of the diffusion of the perception of Merton's paper \cite{merton1968} we classified all 368 journals using nine macro-categories (see Table \ref{table3}).

\subsubsection{Diffusion pattern in the whole journal set}

\begin{figure}[!h]
    \centering
    \includegraphics[width=8cm]{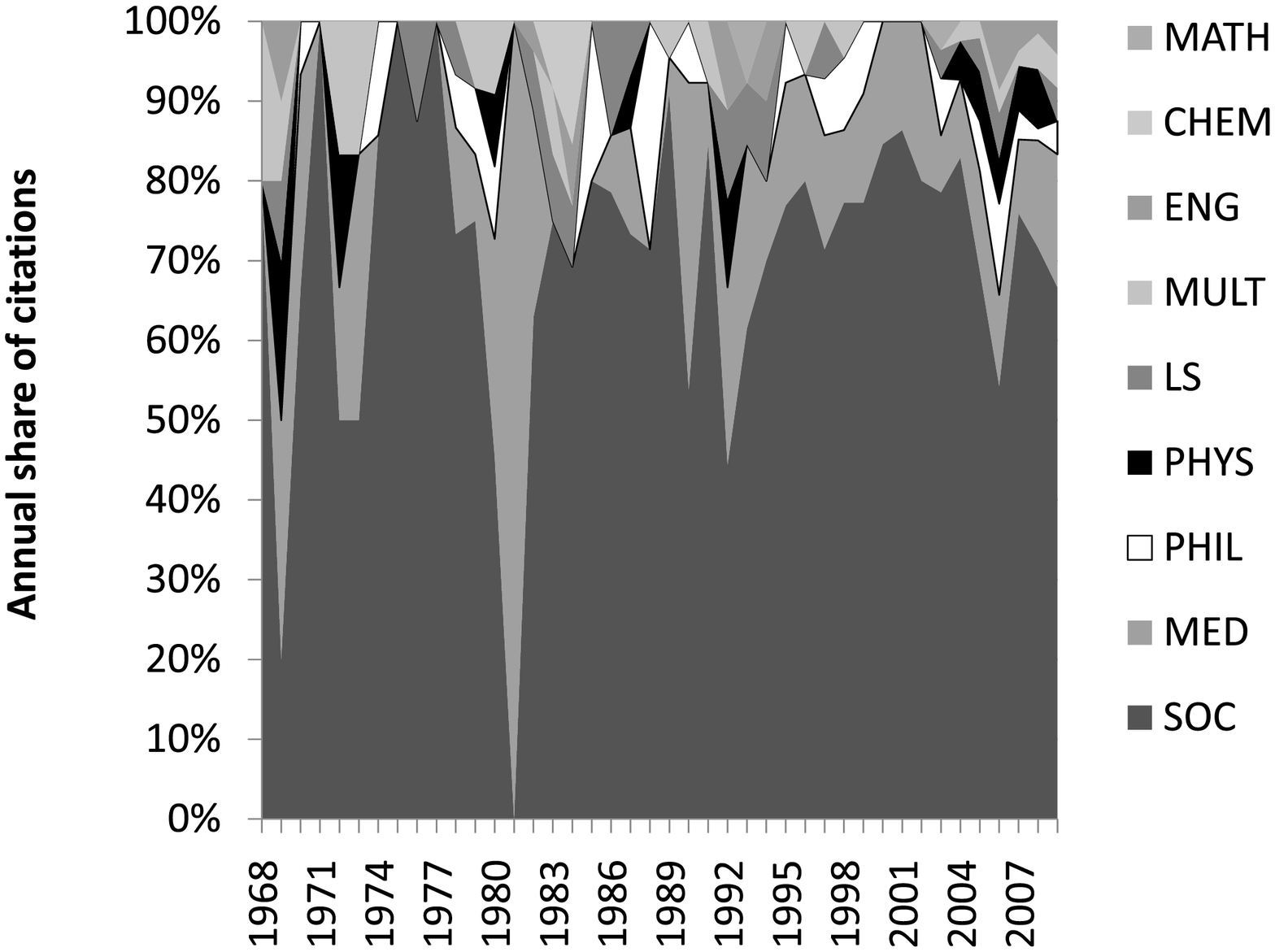}
    \caption{Distribution of papers citing Merton \cite{merton1968} across different disciplines based on a classification of the whole set of journals.}
    \label{andrea7}
\end{figure}

In Fig. \ref{andrea7}, we visualize the annual disciplinary distribution of papers citing Merton's paper \cite{merton1968}. Not surprisingly, the \textit{Social Sciences} (now including \textit{Science and Technology Studies}) dominate the picture. However, each discipline in the natural and social sciences and also a wide variety of journals showed interest in Merton's paper \cite{merton1968}. For instance, in \textit{Physics} the first two citations to Merton's paper \cite{merton1968} appear in 1969, one in the journal ENERGIE NUCLEAIRE (PARIS) with the title \textquotedblleft La Documentation Scientifique et Technique. La Notion de Centre d'Information\textquotedblright\/ \cite{kertesz1969}, and one in the journal PROCEEDINGS OF THE ROYAL SOCIETY OF LONDON SERIES A -- MATHEMATICAL AND PHYSICAL SCIENCES with the title \textquotedblleft Some Problems of Growth and Spread of Science into Developing Countries\textquotedblright.\cite{ziman1969} After 2004, \textit{Physics} seems to pay more systematic attention to Merton's paper of 1968 \cite{merton1968}. In 2004, two papers appeared, one in PHYSICAL REVIEW E (\textquotedblleft Biased Growth Processes and the \textquoteleft rich-get-richer\textquoteright\/ Principle\textquotedblright) \cite{demoura2004} and one in PHYSICS TODAY (\textquotedblleft Could Feynman Have Said This?\textquotedblright) \cite{mermin2004}. The following period of persistent interest in physics journals is due to the emerging specialty \textit{Complex Networks} inside of statistical physics.\cite{scharnhorst2003} Merton's paper \cite{merton1968} has been recognized widely as being important and the fundamental for the understanding of networks of scientific communication and collaboration as nowadays modeled by complex network models.\cite{newman2006}

\begin{figure}[!h]
    \centering
    \includegraphics[width=8cm]{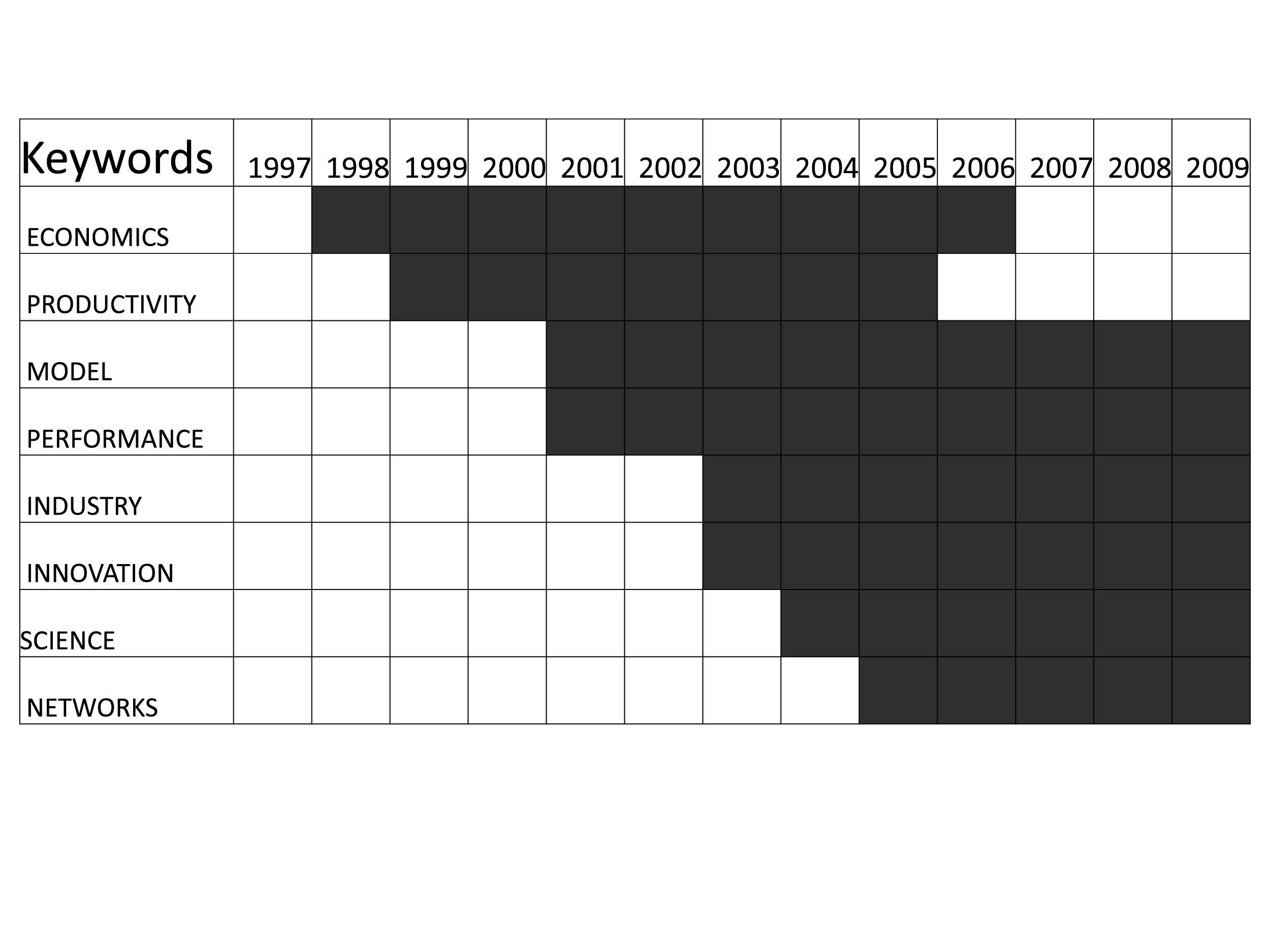}
    \caption{Burst analysis among the keywords -- all years inside of the time period of suddenly increased use are colored}
    \label{andrea9}
\end{figure}

The importance of networks becomes also visible in an experiment we performed with the \textit{Network Workbench}, a tool developed by the group of Katy B\"orner.\cite{nwb2006} This tool allows to import, visualize, and analyze networks, including scientometric data from ISI databases. We experimented with Kleinberg's burst detection algorithm\footnote{According to the documentation of the \textit{NetworkWorkbench}: \textquotedblleft Kleinberg's burst detection algorithm identifies sudden increases in the usage frequency of words. These words may connect to author names, journal names, country names, references, ISI keywords, or terms used in title and/or abstract of a paper. Rather than using plain frequencies of the occurrences of words, the algorithm employs a probabilistic automaton whose states correspond to the frequencies of individual words. State transitions correspond to points in time around which the frequency of the word changes significantly. The algorithm generates a ranked list of the word bursts in the document stream, together with the intervals of time in which they occurred.\textquotedblright \cite{nwb2010}, p. 41.}\cite{kleinberg2003} and applied the algorithm to the keyword field of our dataset. Fig. \ref{andrea9} displays the keywords which are suddenly used more. According to this analysis keyword bursts only occur since mid 1990. Not surprisingly, due to the relevance of Merton's findings for the understanding of the social-behavioral patterns behind complex structures we find \textit{networks} among the  \textquoteleft bursting\textquoteright\/ terms in the period of the emergence of network science \cite{nrc2005,boerner2007} across all disciplines.

\begin{figure}[!h]
    \centering
    \includegraphics[width=10cm]{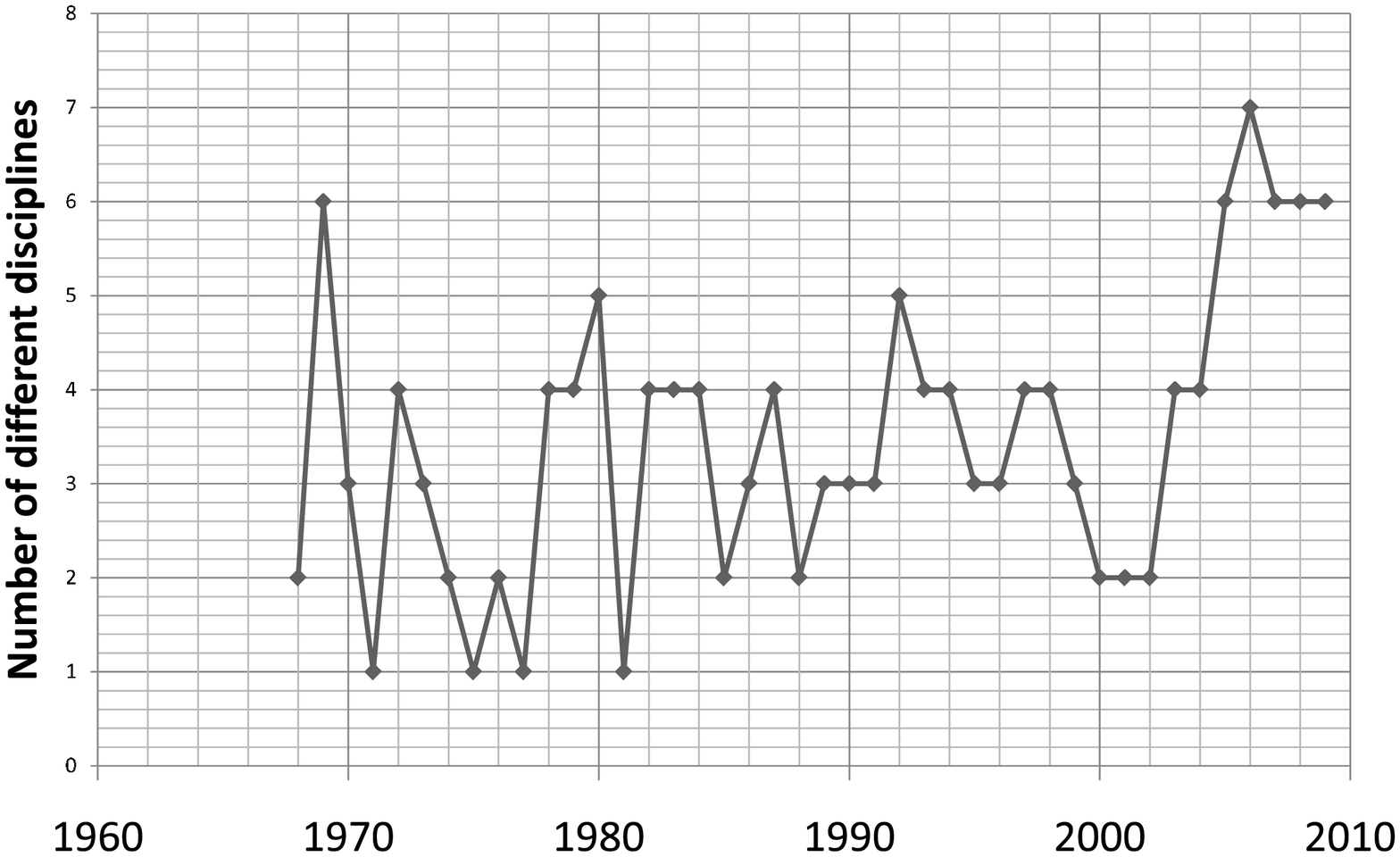}
    \caption{Number of disciplines present each year in the \textquotedblleft recognition sphere\textquotedblright\/ around Merton's paper \cite{merton1968}.}
    \label{andrea8}
\end{figure}

If one looks at the original table of all citing papers and their journals there seems to be an increase both in the number of fields and disciplines present and in the number of citing paper. As we saw in Fig. \ref{andrea7} already, not all disciplines are present at all times. While no clear pattern of diffusion of the perception of Merton's paper \cite{merton1968} among disciplines and no clear transfer path visible are visible, we searched for another indicator to explain the spreading of interest in the work of Merton. However, as shown in Fig. \ref{andrea8} the number of different disciplines present each year shows strong fluctuations which makes it almost impossible to talk about some trends.

\section{Conclusion}
In this paper we explored the possibilities to follow the diffusion or perception of a specific idea using bibliometric approaches. More specifically, we asked why Merton's paper \cite{merton1968} on the Matthew effect in science still lives in the memory of the sciences and still gets cited. We extended historiographic methods as proposed by one of the authors \cite{garfield1973} towards the analysis of the disciplinary spreading of ideas. The visualization of the spreading of ideas on large science maps has been explored recently by Rafols and co-authors.\cite{kiss2010} Our goal was much more modest and much more specific at the same time. We looked at the trace of \textit{one} scholar and, even more narrow, at the trace of one of his publications. We asked what bibliometrics can add to science-historical and biographical research on small scales. 

One access to the disciplinary dimension of the spreading of ideas is to look into the disciplinary origin of papers. Another way would be to follow the disciplinary traces of scholars in an abstract scientific landscape constructed and explored at the same time by scholars traversing it.\cite{scharnhorst2001} Whatever future algorithm will be at hand to map the landscapes of scholarly knowledge on larger and smaller scales, most promising is a combination between \textit{following the actors} (the producers and living \textit{carriers of knowledge}) and their \textit{traces} left. 
\begin{figure}[!h]
    \centering
    \includegraphics[width=10cm]{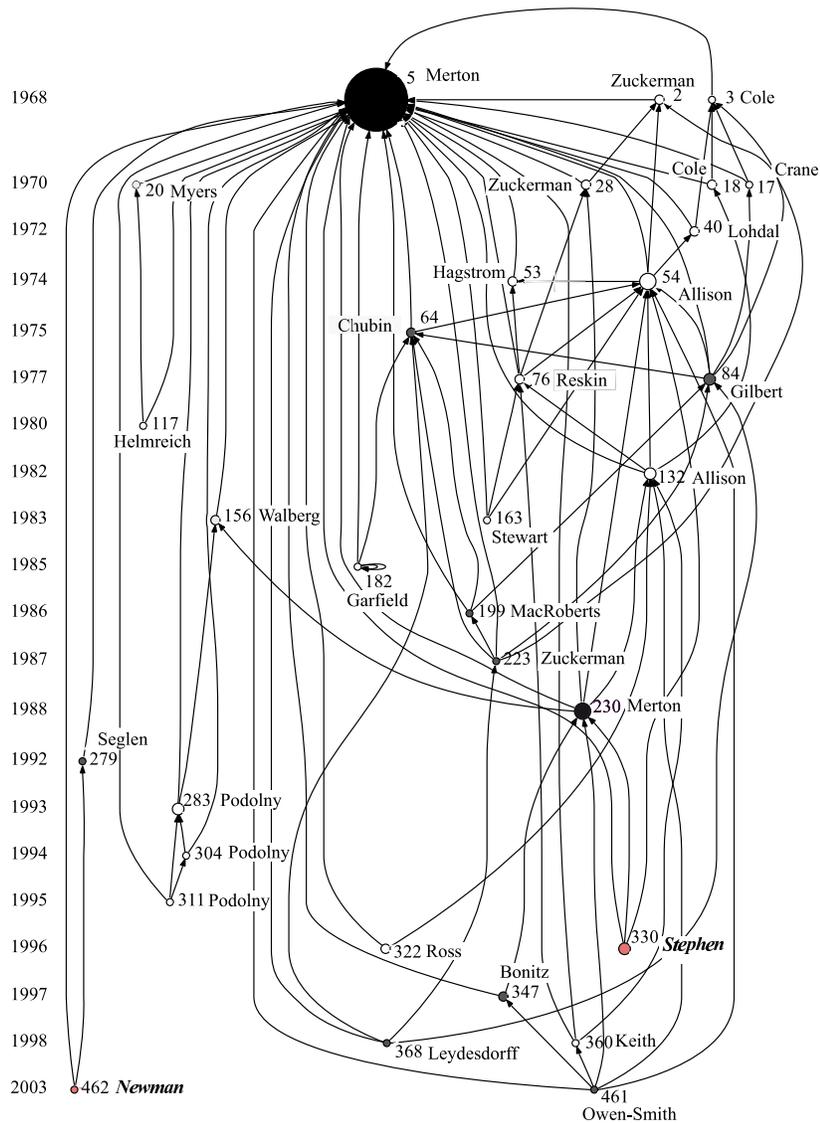}  
    \caption{Historiograph of the citation flows around Merton's paper of 1968 \cite{merton1968}(local G1 graph).}
    \label{andrea10}
\end{figure}
When we tried to visualize this knowledge dynamics in the \textit{HistCite} graph we saw how complex the situation is. Fig. \ref{andrea10} shows the G1 visualization of the citation network of Merton's paper \cite{merton1968}. Let us note once more that in the \textit{HistCite} algorithm the G1 graph selects and visualizes papers according to their importance (citations) \textit{inside} the set of citing papers (local graph). In difference to G1, the G2 graph displays papers according to their importance in the whole database (see Fig. \ref{andrea2}). In Fig. \ref{andrea10}, we have used color codes and first author names in addition to the node numbers. Papers in \textit{Sociology} and \textit{Psychology} jounals appear as white nodes, Merton's papers as black nodes, and papers in \textit{Science and Technology Studies/Information and Documentation Journals} as (dark/light) grey nodes. Not unexpectedly, a look into the specific community structure around Merton's paper \cite{merton1968} reveals almost the same journals which contain most of the citations to its papers.\footnote{A list of the references for all nodes can be found in the Appendix.} With the exception of a paper by Newman (node 426) in a \textit{Physics} journal and a publication by Stephen in an \textit{Economics} journal (node 330) all nodes in this graph either belong to \textit{Sociology} (or \textit{Sociology of Education and Psychology}) or to \textit{Science and Technology Studies}.

The classification exercise inside of the HistCite graph reveals the emergence of new scientific fields (as \textit{Science and Technologies Studies} in the 1970s) and related journals. Node 64 represents the first paper from a \textit{Science and Technologies Studies} journal -- namely SOCIAL STUDIES OF SCIENCE in 1975. Remember that this graph contains only a selection of all citing papers. We have seen the emergence of the new field of \textit{Science and Technology Studies} and the corresponding shift in perception of Merton's paper \cite{merton1968} already in Fig. \ref{andrea6}. Fig. \ref{andrea10} adds concrete \textit{faces} to this shift by visualizing some key papers. We also see that with the time more \textbf{ST*} nodes appear.

At least two mechanisms are important for the diffusion of ideas: researchers which get \textit{infected} by an idea and travelling around taking the idea to new places, and the emergence of new journals which present new channels of communication and are a sign for the formation of new scientifc communities. For instance, in Fig. \ref{andrea10} some authors reappear, even in the narrow selection of G1, and not always they publish in the same scientific field. Irrespectively where published, all articles citing Merton's paper \cite{merton1968} (see Appendix) contribute to a better understanding of the dynamics of the science system and its impact on society. Merton's ideas about (self-)organizing processes inside the science system spread out over different disciplines and over time. Merton's paper \cite{merton1968} has been proven to be a landmark for the study of science. At the same time, Merton's paper is a constitutive element for the formation of a community of researchers interested in science studies whose work forms the basis on which Merton's paper can finally function as a landmark. The diffusion of citations to Merton's key paper \cite{merton1968} across journals, disciplines, and time eventually shows the persistent importantance of the idea for \textit{Social Studies of Science}. This concept is the constant, stable, core knowledge element still floating around, witnessing the longevity and integrating power of scientific ideas.
\subsection{Acknowledgement} Part of this work was supported by COST Action MP0801 \textquotedblleft Physics of Competition and Conflicts\textquotedblright.

\subsection{Appendix}
The following list contains all references displayed in Fig. \ref{andrea10}. For more information please consult http://garfield.library.upenn.edu/histcomp/merton-matthew-I/index-lcs.html.

\begin{description}
\item[\ \ 2] Zuckerman, HA, Patterns of Name Ordering Among Authors of Scientific Papers -- Study of Social Symbolism and Its Ambiguity, AMERICAN JOURNAL OF SOCIOLOGY. 1968; 74 (3): 276-291
\item[\ \ 3] Cole S, Cole JR, Visibility and Structural Bases of Awareness of Scientific Research. AMERICAN SOCIOLOGICAL REVIEW. 1968; 33 (3): 397-413
\item[\ \ 5] Merton RK, Matthew Effect in Science. SCIENCE. 1968; 159 (3810): 56-63
\item[\ 17] Crane D, Academic Marketplace Revisited -- Study of Faculty Mobility Using Cartter Ratings. AMERICAN JOURNAL OF SOCIOLOGY. 1970; 75 (6): 953-964
\item[\ 18] Cole S, Professional Standing and Reception of Scientific
  Discoveries. \hfill\newline AMERICAN JOURNAL OF SOCIOLOGY. 1970; 76 (2): 286-306
\item[\ 20] Myers CR, Journal Citations and Scientific Eminence in Contemporary Psychology. AMERICAN PSYCHOLOGIST. 1970; 25 (11): 1041-1048
\item[\ 28] Zuckerman H, Stratification in American Science. SOCIOLOGICAL INQUIRY. 1970; 40 (2): 235-247
\item[\ 40] Lodahl JB, Gordon G, Structure of Scientific Fields and Functioning of University Graduate Departments. AMERICAN SOCIOLOGICAL REVIEW. 1972; 37 (1): 57-72
\item[\ 53] Hagstrom WO, Competition in Science. AMERICAN SOCIOLOGICAL REVIEW. 1974; 39 (1): 1-18
\item[\ 54] Allison PD, Stewart JA, Productivity Differences Among Scientists -- Evidence for Accumulative Advantage. AMERICAN SOCIOLOGICAL REVIEW. 1974; 39 (4): 596-606
\item[\ 64] Chubin DE, Moitra SD, Content-Analysis of References -- Adjunct or Alternative to Citation Counting. SOCIAL STUDIES OF SCIENCE. 1975; 5 (4): 423-441
\item[\ 76] Reskin BF, Scientific Productivity and Reward Structure of
  Science. \hfill\newline AMERICAN SOCIOLOGICAL REVIEW. 1977; 42 (3): 491-504
\item[\ 84] Gilbert GN, Referencing as Persuasion. SOCIAL STUDIES OF SCIENCE. 1977; 7 (1): 113-122
\item[117] Helmreich RL, Spence JT, Beane WE, Lucker GW, et al., Making it in Academic Psychology -- Demographic and Personality-Correlates of Attainment. JOURNAL OF PERSONALITY AND SOCIAL PSYCHOLOGY. 1980; 39 (5): 896-908
\item[132] Allison PD, Long JS, Krauze TK, Cumulative Advantage and Inequality in Science. AMERICAN SOCIOLOGICAL REVIEW. 1982; 47 (5): 615-625
\item[156] Walberg HJ, Tsai SL, Matthew Effects in Education. AMERICAN EDUCATIONAL RESEARCH JOURNAL. 1983; 20 (3): 359-373
\item[163] Stewart JA, Achievement and Ascriptive Processes in the Recognition of Scientific Articles. SOCIAL FORCES. 1983; 62 (1): 166-189
\item[182] Garfield E, Uses and Misuses of Citation Frequency. CURRENT CONTENTS. 1985; (43): 3-9
\item[199] MacRoberts MH, MacRoberts BR, Quantitative Measures of Communication in Science -- A Study of the Formal Level. SOCIAL STUDIES OF SCIENCE. 1986; 16 (1): 151-172
\item[223] Zuckerman H, Citation Analysis and the Complex Problem of Intellectual Influence. SCIENTOMETRICS. 1987; 12 (5-6): 329-338
\item[230] Merton RK, The Matthew Effect in Science. 2. Cumulative Advantage and the Symbolism of Intellectual Property. ISIS. 1988; 79 (299): 606-623
\item[279] Seglen PO, The Skewness of Science. JOURNAL OF THE AMERICAN SOCIETY FOR INFORMATION SCIENCE. 1992; 43 (9): 628-638
\item[283] Podolny JM, A Status-Based Model of Market Competition. AMERICAN JOURNAL OF SOCIOLOGY. 1993; 98 (4): 829-872
\item[304] Podolny JM, Market Uncertainty and The Social Character of Economic Exchange. ADMINISTRATIVE SCIENCE QUARTERLY. 1994; 39 (3): 458-483
\item[311] Podolny JM, Stuart TE, A Role-Based Ecology of Technological-Change. AMERICAN JOURNAL OF SOCIOLOGY. 1995; 100 (5): 1224-1260
\item[322] Ross CE, Wu CL, Education, Age, and the Cumulative Advantage in Health. JOURNAL OF HEALTH AND SOCIAL BEHAVIOR. 1996; 37 (1): 104-120
\item[330] Stephan PE, The Economics of Science. JOURNAL OF ECONOMIC LITERATURE. 1996; 34 (3): 1199-1235
\item[347] Bonitz M, Bruckner E, Scharnhorst A, Characteristics and Impact of the Matthew Effect for Countries. SCIENTOMETRICS. 1997; 40 (3): 407-422
\item[360] Keith B, Babchuk N, The Quest for Institutional Recognition: A Longitudinal Analysis of Scholarly Productivity and Academic Prestige among Sociology Departments. SOCIAL FORCES. 1998; 76 (4): 1495-1533
\item[368] Leydesdorff L, Theories of Citation? SCIENTOMETRICS. 1998; 43 (1): 5-25
\item[461] Owen-Smith J, From Separate Systems to a Hybrid Order: Accumulative Advantage Across Public and Private Science at Research One Universities. RESEARCH POLICY. 2003; 32 (6): 1081-1104
\item[462] Newman MEJ, The Structure and Function of Complex Networks. SIAM REVIEW. 2003 JUN; 45 (2): 167-256
\end{description}


\begin{thebibliography}{99}

\bibitem{adamic1999} Adamic, L.A., \textit{Zipf, Power-laws, and Pareto -- a
  Ranking Tutorial}. Working paper 1999, Available at 
\hfill\newline http://www.hpl.hp.com/research/idl/papers/ranking/

\bibitem{debellis2009} De Bellis, N., \textit{Bibliometrics and Citation Analysis. From the Science Citation Index to Cybermetrics}. Lanham: The Scarecrow Press, 2009.

\bibitem{bernal1939} Bernal, J.D., \textit{The Social Function of Science}. London: George Routledge \& Sons Ltd., 1939.

\bibitem{bonitz1997} Bonitz, M., \textit{The Scientific Talents of Nations}. Libri \textbf{47}(4)(1997)206-213.

\bibitem{bonitz1999} Bonitz, M., Bruckner, E., Scharnhorst, A., \textit{The
  Matthew Index -- Concentration Patterns and Matthew Core
  Journals}. Scientometrics \hfill\newline \textbf{44}(3)(1999)361-378.

\bibitem{boerner2004} B\"{o}rner, K., Maru, J., Goldstone, R., \textit{The Simultaneous Evolution of Author and Paper Networks}. Proceedings of the National Academy of Sciences of the United States of America, \textbf{101}(Suppl. 1)(2004)5266-5273.

\bibitem{boerner2007} B\"{o}rner, K.,  Sanyal, S., Vespignani, A., \textit{Network Science}. In: Cronin, B. (Ed.), Annual Review of Information Science and Technology, Vol. 41, pp. 537-607, Chapter 12, Medford, NJ: Information Today, Inc./American Society for Information Science and Technology, 2007.

\bibitem{boyack2005} Boyack, K.W., Klavans, R., B\"{o}rner, K., \textit{Mapping the Backbone of Science}. Scientometrics \textbf{64}(3)(2005)351-374.

\bibitem{bruckner1990} Bruckner, E., Ebeling, W., Scharnhorst, A., \textit{The Application of Evolution Models in Scientometrics}. Scientometrics \textbf{18}(1-2)(1990) 21-41.

\bibitem{bruckner1996} Bruckner, E. Ebeling, W., Jim\'enez-Monta\~no, M.A., Scharnhorst, A., \textit{Nonlinear Effects of Substitution -- an Evolutionary Approach}. Journal of Evolutionary Economics \textbf{6}(1)(1996)1-30. 

\bibitem{cronin2000} Cronin, B., Atkins, H. (Eds.), \textit{The Web of Knowledge}. A Festschrift in Honor of Eugene Garfield. Medford, New Jersey: Information Today, Inc., 2000.

\bibitem{dobrov1971} Dobrov, G.M., \textit{Potential der Wissenschaft}. Berlin: Akademie-Verlag, 1971.

\bibitem{elkana1978} Elkana, Y. Lederberg, J., Merton, R.K., Thackray, A., Zuckerman, H. (Eds.). \textit{Toward a Metric of Science: The Advent of Science Indicators}. New York: John Wiley \& Sons,  1978.

\bibitem{fronczak2007} Fronczak, P., Fronczak, A., Holyst, J.A., \textit{Analysis of Scientific Productivity Using Maximum Entropy Principle and Fluctuation-Dissipation Theorem}. Physical Review E \textbf{75}(2)(2007)026103, 9 pages.


\bibitem{garfield1973} Garfield, E., \textit{Historiographs, Librarianship, and the History of Science}. In: Rawski, C.H. (Ed.), Toward a Theory of
Librarianship: Papers in Honor of Jesse Hauk Shera, N. J.: Scarecrow Press, 1973, p. 380-402. Reprinted in: Garfield, E., Essays of an Information Scientist, Vol. 2, 1974-76, pp. 136-150; and Current Contents, \textbf{\#38}, September 18, 1974. Available at http://www.garfield.library.upenn.edu/essays/v2p136y1974-76.pdf 

\bibitem{garfield1986} Garfield, E., \textit{Do Nobel Prize Winners Write
  Citation Classics?} Current Contents, \textbf{\#23}(1986)3-8; reprinted in:
  Garfield, E., Essays of an Information Scientist, Vol. 9, 1986, p. 182;
  Available at \hfill\newline 
http://www.garfield.library.upenn.edu/essays/v9p182y1986.pdf
 
\bibitem{garfield2004a} Garfield, E., \textit{The Intended Consequences of Robert K. Merton}. Scientometrics \textbf{60}(1)(2004)51-61.

\bibitem{garfield2004b} Garfield, E. \textit{The Unintended and Unanticipated Consequences of Robert K. Merton}. Social Studies of Science \textbf{34}(6)(2004)845-854.

\bibitem{garfield2003} Garfield, E., Pudovkin, A.I., Istomin, V.I., \textit{Mapping the Output of Topical Searches in the Web of Knowledge and the Case of Watson-Crick}. Information Technology and Libraries \textbf{22}(4)(2003)183-187.  

\bibitem{glaenzel2003} Gl\"anzel, W., \textit{Bibliometrics as a Research Field: A Course on Theory and Application of Bibliometric Indicators}. Courses Handout, On-line Source, Available at http://www.norslis.net/2004/Bib\_Module\_KUL.pdf (accessed on February 2, 2010), 2003.

\bibitem{hellsten2008} Hellsten, I., Lambiotte, R., Scharnhorst, A., Ausloos, M., \textit{Self-citations, Co-authorships and Keywords: A New Approach to Scientists' Field Mobility?} Scientometrics \textbf{72}(3)(2008)469-486.

\bibitem{kertesz1969} Kertesz, F., \textit{La Documentation Scientifique et Technique. La Notion de Centre d'Information}. Energie Nucl\'eaire (Paris) \textbf{11}(6)(1969)366-371.

\bibitem{kiss2010} Kiss, I.Z., Broom, M., Craze. P.G., Rafols, I., \textit{Can Epidemic Models Describe the Diffusion of Topics Across Disciplines?} Journal of Informetrics \textbf{4}(1)(2010)74-82. 
 
\bibitem{kleinberg2003} Kleinberg, J., \textit{Bursty and Hierarchical Structure in Streams}. Data Mining and Knowledge Discovery \textbf{7}(4)(2003)373-397.

\bibitem{leydesdorff2010} Leydesdorff, L., \textit{What Can Heterogeneity Add to the Scientometric Map? Steps Towards Algorithmic Historiography}. In: Akrich,M.,  Barthe,Y.,  Muniesa, F., Mustar, P. (Eds.), Festschrift for Michel Callon's 65th Birthday. Paris: \'Ecole Nationale Sup\'erieure des Mines, 2010 (in press). Preprint available at arxiv.org (arXiv:1002.0532v1). 

\bibitem{lucio2008} Lucio-Arias, D., Leydesdorff, L., \textit{Main-Path Analysis and Path-dependent Transitions in HistCite\texttrademark-based Historiograms}. Journal of the American Society for Information Science and Technology \textbf{59}(12)(2008)1948-1962.

\bibitem{mermin2004} Mermin, N.D., \textit{Could Feynman Have Said This?}
  Physics Today \hfill\newline \textbf{57}(5)(2004)10.

\bibitem{merton1949} Merton, R.K., \textit{Social Theory and Social Structure}. New York: Free Press, 1949.

\bibitem{merton1957} Merton, R.K., \textit{Priorities in Scientific Discovery: A Chapter in the Sociology of Science}. American Sociological Review \textbf{22}(6)(1957)635-659.

\bibitem{merton1968} Merton, R.K., \textit{The Matthew Effect in Science: The Reward and Communication Systems of Science are Considered}. Science \textbf{159}(No. 3810)(1968)56-63.

\bibitem{merton1980} Merton, R.K., \textit{This Week's Citation Classics: Merton, R. K., Social Theory and Social Structure. New York: Free Press, 1949. 423 pp.
[Columbia University, New York, NY]} Current Contents \textbf{21}(1980)285, Available at www.garfield.library.upenn.edu/classics1980/A1980JS04600001.pdf

\bibitem{merton1988} Merton, R.K., \textit{The Matthew Effect in Science, II: Cumulative Advantage and the Symbolism of Intellectual Property}. ISIS \textbf{79}(1988)606-623. 

\bibitem{merton1995} Merton, R.K., \textit{The Thomas Theorem and The Matthew Effect}. Social Forces \textbf{74}(2)(1995)379-424.

\bibitem{demoura2004} De Moura, A.P.S., \textit{Biased Growth Processes and the \textquoteleft Rich-Get-Richer\textquoteright\/ Principle}. Physical Review E \textbf{69}(5)(2004)056116, 5 pages.

\bibitem{nalimov1969} Nalimov, V.V., Mul'chenko, Z.M., \textit{Naukometriya}. Moscow: Nauka, 1969.

\bibitem{nrc2005} National Research Council. Committee on Network Science for Future Army Applications. \textit{Network Science}. Washington: The National Acade\-mies Press, 2005.

\bibitem{nwb2006} NWB Team. \textit{Network Workbench Tool}. Indiana University, Northeastern University, and University of Michigan, http://nwb.slis.indiana.edu, 2006.

\bibitem{nwb2010}NetworkWorkbench. 2010. \textit{User manual 1.0.0.}
  (Available at 
\hfill\newline http://nwb.slis.indiana.edu/Docs/NWBTool-Manual.pdf, accessed February 23, 2010).

\bibitem{newman2004} Newman, M., \textit{Coauthorship Networks and Patterns of Scientific Collaboration}. Proceedings of the National Academy of Sciences of the United States of America \textbf{101}(Suppl. 1)(2004)5200-5205.

\bibitem{newman2006} Newman, M., Barab\'asi, A.L., Watts, D.J., \textit{The Structure and Dynamics of Networks}. Princeton, Oxford: Princeton University Press, 2006.

\bibitem{price1965} Price, D.J. de Solla, \textit{Networks of Scientific Papers: The Pattern of Bibliographic References Indicates the Nature of the Scientific Research Front}. Science \textbf{149}(No. 3683)(1965)510-515.

\bibitem{price1976} Price, D.J. de Solla, \textit{General Theory of Bibliometric and Other Cumulative Advantage Processes}. Journal of the American Society for Information Science \textbf{27}(5-6)(1976)292-306.

\bibitem{prigogine1986} Prigogine, I., Stengers, I., \textit{Order Out of Chaos: Man's New Dialogue with Nature}. London: Fontana Paperbacks, 1986.

\bibitem{scharnhorst2001} Scharnhorst, A., \textit{Constructing Knowledge Landscapes within the Framework of Geometrically Oriented Evolutionary Theories}. In: Matthies, M., Malchow, H., Kriz, J.(Eds.), Integrative Systems Approaches to Natural and Social Sciences -- Systems Science 2000. Berlin: Springer, 2001, pp. 505-515.

\bibitem{scharnhorst2003} Scharnhorst, A., \textit{Complex Networks and the Web: Insights from Nonlinear Physics}. Journal of Computer-Mediated Communication \textbf{8}(4)2003, http://jcmc.indiana.edu/vol8/issue4/scharnhorst.html.

\bibitem{steiner1989} Steiner, H. (Ed.), \textit{J. D. Bernal's The Social Function of Science: 1939 -- 1989}. Berlin: Akademie-Verlag, 1989.

\bibitem{suarez1995} Su\'arez, F.F., Utterback, J.M., \textit{Dominant Designs and the Survival of Firms}. Strategic Management Journal \textbf{16}(6)(1995)415-430.

\bibitem{vlachy1981} Vlachy, J., \textit{Mobility in Physics}. Czechoslovak Journal of Physics\hfill\newline \textbf{31}(6)(1981)669-674.

\bibitem{vlachy1983} Vlachy, J., \textit{Tracing Innovative Papers in Physics by Successive Citation -- Concept and Exemplars}. Czechoslovak Journal of Physics \textbf{B33}(1983)841-844.

\bibitem{wouters1999} Wouters, P., \textit{The Citation Culture}. Amsterdam:
  University of Amsterdam Amsterdam, PhD Thesis (unpublished),
  1999.\hfill\newline (Available
  http://garfield.library.upenn.edu/wouters/wouters.pdf,
\hfill\newline accessed February 2, 2010)

\bibitem{ziman1969} Ziman, J.M., \textit{Some Problems of Growth and Spread of Science into Developing Countries}. Proceedings of The Royal Society of London Series A -- Mathematical and Physical Sciences \textbf{311}(No. 1506)(1969)349-369.

\bibitem{zuckerman2010} Zuckerman, H., \textit{The Matthew Effect Writ Large and Larger: A Study in Sociological Semantics}. In: Elkana, Y. (Ed.), Concepts and Social Order:  Essays on the Works of Robert K. Merton. Budapest: Central University Press, 2010 (in press).

\end{thebibliography}
\end{document}